# Can Commercial LLMs Be Parliamentary Political Companions?

*Comparing LLM Reasoning Against Romanian Legislative Expuneri de Motive*

Iulian Lucau[1]     Adelin-George Voicu[2]

[1]*GuidingAI, Sønderborg, Denmark*

[2]*GuidingAI, Sønderborg, Denmark*

**Abstract**

This paper evaluates whether commercial large language models (LLMs) can function as reliable political advisory tools by comparing their outputs against official legislative reasoning. Using a dataset of 15 Romanian Senate law proposals paired with their official explanatory memoranda (*expuneri de motive*), we test six LLMs spanning three provider families and multiple capability tiers: GPT-5-mini, GPT-5-chat (OpenAI), Claude Haiku 4.5 (Anthropic), and Llama 4 Maverick, Llama 3.3 70B, and Llama 3.1 8B (Meta). Each model generates predicted rationales evaluated through a dual framework combining LLM-as-Judge semantic scoring and programmatic text similarity metrics. We frame the LLM-politician relationship through principal-agent theory and bounded rationality, conceptualizing the legislator as a principal delegating advisory tasks to a boundedly rational agent under structural information asymmetry. Results reveal a sharp two-tier structure: frontier models (Claude Haiku 4.5, GPT-5-chat, GPT-5-mini) achieve statistically indistinguishable semantic closeness scores above 4.6/5.0, while open-weight models cluster a full tier below (Cohen's d > 1.4). However, all models exhibit task-dependent confabulation, performing well on standardized legislative templates (e.g., EU directive transpositions) but generating plausible yet unfounded reasoning for politically idiosyncratic proposals. We introduce the concept of cascading bounded rationality to describe how failures compound across bounded principals, agents, and evaluators, and argue that the operative risk for legislators is not stable ideological bias but contextual ignorance shaped by training data coverage.

# 1 Introduction

Political decision-making operates under conditions of bounded rationality (Simon, 1990). Legislators face an ever-expanding volume of policy proposals, legal amendments, and constituent demands while constrained by limited time and finite cognitive resources. To compensate, they rely on advisory ecosystems: staff, party analysts, lobbyists, and think tanks, that filter, synthesize, and frame information for action. The rapid proliferation of commercial large language models (LLMs) has introduced a new category of potential advisory tool: AI systems capable of analyzing legislation, drafting policy arguments, and generating strategic recommendations on demand. Whether these systems can function as reliable political companions is an open empirical question with significant implications for governance.

A growing body of research has examined LLMs in political contexts, establishing that most commercial models exhibit measurable left-leaning political bias on orientation tests (Rozado, 2024; Yang et al., 2024). Studies have demonstrated that LLMs can simulate voter behavior (Yu et al., 2025; Zhou et al., 2025), predict legislative roll-call votes (Li et al., 2025), and persuade citizens on policy issues with effects rivaling human-crafted materials (Hackenburg et al., 2025; Bai et al., 2025). However, this literature overwhelmingly treats LLMs as objects of study or potential threats to democratic processes. The practitioner's question, whether a politician or legislative staffer can rely on these tools for day-to-day advisory work, remains largely unaddressed. Furthermore, existing bias evaluations rely on abstract political quizzes disconnected from actual legislative reasoning, and nearly all empirical work is confined to English-language, U.S.-centric political contexts.

This paper addresses these gaps by evaluating six LLMs spanning three provider families as political advisory agents using real legislative data from the Romanian Senate and Deputy Chamber. We test GPT-5-mini and GPT-5-chat (OpenAI), Claude Haiku 4.5 (Anthropic), and Llama 4 Maverick, Llama 3.3 70B, and Llama 3.1 8B (Meta). This selection encompasses frontier closed-source models and open-weight models ranging from 8 billion to 400 billion parameters, enabling analysis of how provider family, model scale, architectural choices (dense versus mixture-of-experts), and reasoning capabilities (chain-of-thought versus standard) affect political advisory performance. We frame the LLM-politician relationship through principal-agent theory (Jensen & Meckling, 1976): the politician (principal) delegates analytical tasks to the LLM (agent), creating information asymmetry, potential incentive misalignment through embedded training biases, and an adverse selection problem when choosing among competing models. To operationalize this framework, we feed actual Romanian Senate law proposals to each LLM and evaluate the resulting outputs against official explanatory memoranda (*expuneri de motive*) as ground truth, employing a dual evaluation design combining LLM-as-Judge semantic scoring (Zheng et al., 2023) and programmatic text similarity metrics.

The main research question asks: how reliable are commercial LLMs as political advisory tools when evaluated against official legislative reasoning from the Romanian Senate? RQ1 examines how models across different provider families and capability tiers compare in reasoning alignment when analyzing real Romanian legislative proposals. RQ2 investigates whether the ideological bias documented in abstract settings persists when models engage with real law proposals, or whether alternative failure modes, such as task-dependent confabulation, prove more consequential in applied legislative advisory contexts.

This work aims to make three contributions. First, it provides the first practitioner-oriented empirical evaluation of commercial LLMs as political advisory tools using real parliamentary data, testing six models across three provider families. Second, it extends the political bias literature beyond abstract quizzes to applied legislative advisory tasks, revealing that task-dependent confabulation shaped by training data coverage poses a greater risk than stable ideological lean. Third, it introduces evidence from a non-Anglophone EU member state, Romania, where LLM training data representation is substantially thinner than for dominant-language political systems, and proposes the concept of cascading bounded rationality to characterize how failures compound across bounded principals, agents, and evaluators.

## 2 Related Work

### 2.1 LLMs and Political Bias

The detection of political bias in LLMs has become a growing research area. Rozado (2024) administered eleven political orientation tests to twenty-four LLMs and found most generated left-of-center responses; a subsequent study combined four complementary methodologies to produce a unified bias ranking (Rozado, 2025). Yang et al. (2024) scaled this inquiry to forty-three LLMs, confirming a Democratic-leaning preference in seventy-six percent of models on the 2024 U.S. presidential race. Choudhary (2024) reported that ChatGPT-4 and Claude exhibited liberal tendencies while Gemini adopted centrist stances. A language-dependent dimension was identified by Yuksel et al. (2025), who found that political orientation shifted significantly depending on query language. More recently, Promptfoo (2025) benchmarked frontier models on 2,500 political statements and found all scored left of center, with Claude Opus 4 closest to neutral. A persistent limitation is methodological: bias is measured through abstract instruments designed for human test-takers, not through applied tasks reflecting actual practitioner interaction.

### 2.2 LLMs as Political Agents and Simulators

Parallel research has examined LLMs as agents simulating political behavior. Li et al. (2025) proposed the Political Actor Agent framework for predicting U.S. Congressional roll-call votes. Kreutner et al. (2025) simulated European Parliament voting behavior, achieving a weighted F1 of 0.793 with zero-shot prompting. Yu et al. (2025) developed a multi-step reasoning framework to simulate U.S. elections at scale, while Zhou et al. (2025) introduced FlockVote, replicating the 2024 election outcome while exposing pervasive instability across GPT-4o, Claude-3.5-sonnet, and Gemini-1.5-Pro. Moghimifar et al. (2024) modeled coalition negotiations as hierarchical Markov decision processes. These studies demonstrate that LLMs can approximate political reasoning under structured conditions, but outputs are sensitive to prompt phrasing, exhibit non-deterministic instability, and default to progressive positions when persona controls are removed.

### 2.3 LLM Persuasion and Decision-Support in Politics

The persuasive capacity of LLMs has received increasing attention. Hackenburg et al. (2025) demonstrated in *Science* that conversational LLM interactions shift voters' political attitudes. Potter et al. (2024) showed that exposure to biased LLM agents changes real voters' opinions, establishing downstream behavioral consequences. Chen et al. (2026) benchmarked seven frontier LLMs in experiments with over nineteen thousand participants, finding all models outperformed actual campaign advertisements, with persuasion asymmetrically stronger toward Democratic positions. On the decision-support side, Loffredo & Yun (2025) demonstrated that LLM agents can perform labor-intensive legislative research tasks, while Bandaru et al. (2025) found that multi-agent LLM debates converge toward left-leaning positions even when assigned opposing identities. These findings suggest that while LLMs possess capabilities relevant to political advisory work, their reliability as decision-support tools cannot be assumed and requires domain-specific validation.

## 2.4 LLM-as-Judge Methodology

The evaluation of open-ended LLM outputs has increasingly relied on LLMs themselves as evaluators. Zheng et al. (2023) introduced the MT-Bench and Chatbot Arena frameworks, establishing that strong LLMs can serve as scalable alternatives to human judges. Subsequent work identified systematic self-preference bias, where models score their own outputs higher (Panickssery et al., 2024), motivating cross-model evaluation designs. The LLM-as-Judge approach has been applied to coding benchmarks and creative writing assessment but has not been used to evaluate political advisory quality against real parliamentary ground truth. The present study adapts this methodology for legislative advisory evaluation.

# 3 Theoretical Framework

## 3.1 Bounded Rationality and the Advisory Imperative

Simon's (1990) model of bounded rationality posits that decision-makers cannot optimize because they lack the cognitive capacity to process all available information. Instead, they satisfy, selecting options meeting a threshold of acceptability. Kahneman (2011) further decomposed decision-making into fast, intuitive System 1 and slow, deliberative System 2 processes, arguing that cognitive constraints push decision-makers toward heuristic-driven shortcuts susceptible to systematic biases. Gigerenzer and Gaissmaier (2011) challenged this by arguing that in uncertain environments, simple decision rules can outperform complex optimization.

This debate is directly relevant to LLM advisory. If an LLM functions as a provider of fast but potentially biased analysis, it may replicate the System 1 shortcuts Kahneman warns against, fluent outputs masking factual gaps. Alternatively, if the LLM surfaces considerations the legislator would not have encountered independently, it may function as a bounded rationality mitigator. The question is not whether LLMs produce perfect advice, no advisory source does, but whether they reduce or amplify the information deficit under which legislators operate.

### 3.2 The Principal-Agent Problem in LLM-Mediated Advisory

Jensen and Meckling (1976) formalized the principal-agent relationship as one in which a principal delegates tasks to an agent whose interests may diverge. Agency costs arise from monitoring costs, bonding costs, and residual loss. Eisenhardt (1989) identified two core problems: adverse selection (the principal cannot assess agent quality prior to engagement) and moral hazard (the principal cannot observe whether the agent acts faithfully once engaged). Both are mediated by information asymmetry.

Applying this framework to the LLM-politician relationship surfaces different tensions. With a human advisor, the politician can evaluate track record, affiliations, and ideological commitments, the information asymmetry is bounded by social legibility. With an LLM, the asymmetry deepens categorically: the politician cannot inspect training data composition, RLHF objectives, or the alignment procedures shaping outputs. The biases documented by Rozado (2024) and Yang et al. (2024) are not disclosed to users and are unstable across versions, creating information asymmetry that is structurally opaque. An LLM's misalignment is embedded in statistical regularities absorbed from training corpora, neither strategic nor correctable through reputational incentives. The adverse selection problem is directly addressed by this study: a politician choosing among competing models faces a market with no domain-specific quality signals.

### 3.3 Integration: LLM-as-Judge as Monitoring Mechanism

Principal-agent theory prescribes monitoring as the primary mechanism for mitigating agency costs (Eisenhardt, 1989). The LLM-as-Judge methodology (Zheng et al., 2023) functions as an automated monitoring mechanism: a third-party evaluator assessing output quality on behalf of a principal who lacks independent evaluation capacity. The cross-model design mirrors independent auditing logic, justified by documented self-preference bias (Panickssery et al., 2024). However, this introduces a second-order agency problem: the judge is itself a boundedly rational agent. If all commercial LLMs share similar training distributions, the monitor may be blind to the most pervasive failures.

The integration produces a layered framework: bounded rationality establishes demand (legislators need advisory support); principal-agent theory identifies risks (information asymmetry, embedded bias, adverse selection); and the LLM-as-Judge protocol operationalizes monitoring while bounded rationality reminds us the monitor itself operates under constraints. This structure, bounded agents evaluating bounded agents on behalf of bounded principals, is the theoretical foundation for the empirical evaluation.

## 4 Methodology

Our methodology follows a three-stage pipeline: (1) data preparation, where legislative change documents are paired with their official explanatory rationales; (2) generation, where each law change is sent to multiple LLM endpoints; and (3) evaluation, where we assess output quality using both LLM-as-Judge semantic scoring and programmatic text similarity metrics.

### 4.1 Dataset

We constructed a dataset of 15 Romanian legislative changes paired with their official *expuneri de motive*. Each sample consists of a structured description of the law change (modified law reference, article-level amendments, comparison table, and summary of new elements) and the official explanatory rationale serving as ground truth. The law changes span labor law, criminal law, and administrative procedures. Document lengths range from approximately 2,000 to 35,000 characters for ground truth rationales. All documents are in Romanian, extracted from the Chamber of Deputies and Senate websites and converted to structured plain text.

### 4.2 Models Under Evaluation

We evaluate six LLMs spanning three provider families and multiple capability tiers, selected to test how model scale, architecture, and training methodology affect legal reasoning performance in a non-English language. All models are accessed through OpenRouter API endpoints. Table 1 summarizes the models and selection rationale.

**Table 1:** Models under evaluation.

| Model | Provider | Parameters | Tier | Selection Rationale |
|---|---|---|---|---|
| GPT-5 Mini | OpenAI | Undisclosed | Frontier | Reasoning model; chain-of-thought for legal tasks |
| GPT-5 Chat | OpenAI | Undisclosed | Frontier | Non-reasoning variant; isolates reasoning effect |
| Claude Haiku 4.5 | Anthropic | Undisclosed | Frontier | Efficient frontier; strong multilingual capability |
| Llama 4 Maverick | Meta | 17B/400B MoE | Open-weight | 128-expert MoE; best open-weight in class |
| Llama 3.3 70B | Meta | 70B dense | Open-weight | Established dense baseline |
| Llama 3.1 8B | Meta | 8B dense | Open-weight | Small model lower bound |

The selection includes two GPT-5 variants (reasoning versus non-reasoning) to isolate chain-of-thought effects, frontier models from two providers to test cross-provider consistency, and three Llama models spanning an order-of-magnitude parameter range (8B to 400B MoE) to evaluate scaling effects within the open-weight family.

### 4.3 Generation Procedure

Each law change is sent to each model using a fixed prompt instructing the model to act as a Romanian legislative expert and generate the most likely *exposé des motifs*. The prompt specifies Romanian-language output addressing legal, social, economic, and administrative justifications. The identical prompt is used across all models. Temperature is set to 0.3 with a maximum of 8,192 output tokens; for GPT-5 family models, which do not support the temperature parameter, default sampling is used. Each model-input combination is evaluated across 5 independent runs, yielding 75 traces per model and 450 total traces.

### 4.4 Evaluation Framework

We employ a dual evaluation framework combining semantic assessment via LLM-as-Judge with programmatic text similarity metrics.

**LLM-as-Judge Evaluation.** We use MLflow's custom_prompt_judge to evaluate generated rationales against ground truth across three dimensions, each on a 5-point scale (1=poor to 5=excellent): *Argument Coverage* (whether key arguments from the official memorandum are captured); *Factual Alignment* (consistency of specific facts, statistics, law references, and institutional names); and *Exposé des Motifs Closeness* (holistic semantic similarity considering argument coverage, reasoning depth, factual consistency, and structural organization).

**Programmatic Metrics.** We compute nine metrics spanning four families: n-gram overlap (ROUGE-1, ROUGE-2, ROUGE-L F1); token-set metrics (Jaccard similarity, trigram novelty); embedding-based metrics (cosine similarity via paraphrase-multilingual-MiniLM-L12-v2, TF-IDF cosine); and domain-specific metrics (legal entity overlap via pattern matching of article numbers, law citations, and dates; length ratio). The inclusion of domain-specific metrics is motivated by the observation that legal text quality depends heavily on referencing correct articles and provisions, information that generic similarity metrics may miss.

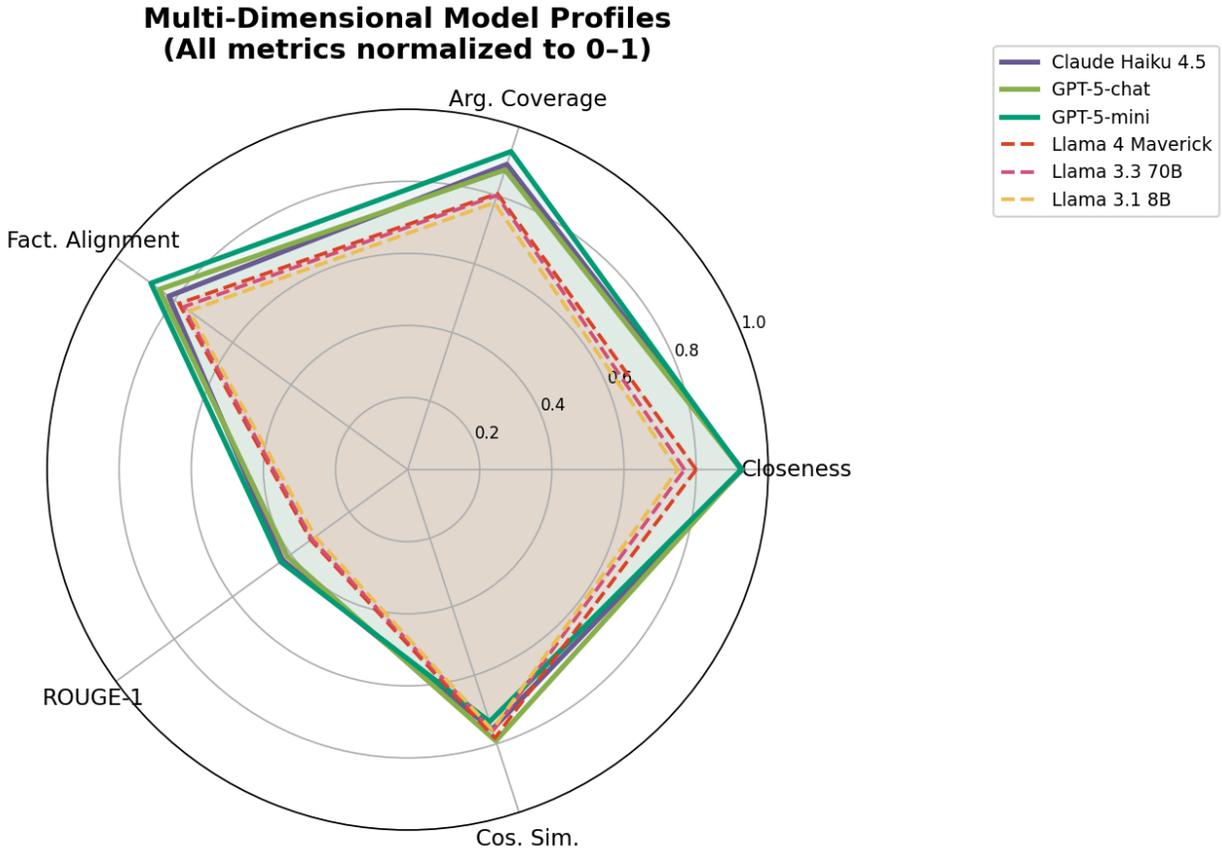

Figure 0: Multi-dimensional model profiles (all metrics normalized to 0–1). Solid lines: Tier 1 (frontier); dashed: Tier 2 (open-weight).

## 5 Results

We evaluated six LLMs across 15 Romanian legislative changes with five runs per model (N = 75 per model, 450 total). Each output was scored by three LLM-as-Judge dimensions and four key programmatic metrics.

### 5.1 Overall Performance: A Clear Two-Tier Structure

The most salient finding is a sharp two-tier separation. Three models, Claude Haiku 4.5, GPT-5-chat, and GPT-5-mini, form a statistically indistinguishable top tier on semantic closeness (means: 4.64, 4.63, 4.63; all pairwise Mann–Whitney $p > 0.86$). The remaining three cluster in a second tier (means: 4.00, 3.84, 3.75). Between-tier differences are highly significant (U = 42,006, $p < 0.001$) with very large effect sizes (Cohen's $d = 1.45$–$1.83$). The between-tier gap dwarfs within-tier differences, suggesting that model family matters far more than size or configuration within a family for this task.

**Table 2:** Mean (±SD) for all metrics by model. Judge metrics on 1–5 scale; ROUGE-1 and cosine similarity on 0–1.

| Model | Closeness | Arg. Cov. | Fact. Align. | ROUGE-1 | Cos. Sim. |
|---|---|---|---|---|---|
| Claude Haiku 4.5 | 4.64 (±0.48) | 4.45 (±0.50) | 4.09 (±0.57) | 0.428 | 0.758 |
| GPT-5-chat | 4.63 (±0.49) | 4.37 (±0.51) | 4.25 (±0.57) | 0.412 | 0.792 |
| GPT-5-mini | 4.63 (±0.49) | 4.64 (±0.48) | 4.40 (±0.49) | 0.435 | 0.735 |
| Llama 4 Maverick | 4.00 (±0.37) | 4.03 (±0.28) | 3.92 (±0.49) | 0.332 | 0.783 |
| Llama 3.3 70B | 3.84 (±0.40) | 4.00 (±0.16) | 3.84 (±0.40) | 0.328 | 0.763 |
| Llama 3.1 8B | 3.75 (±0.50) | 3.89 (±0.31) | 3.75 (±0.47) | 0.317 | 0.760 |

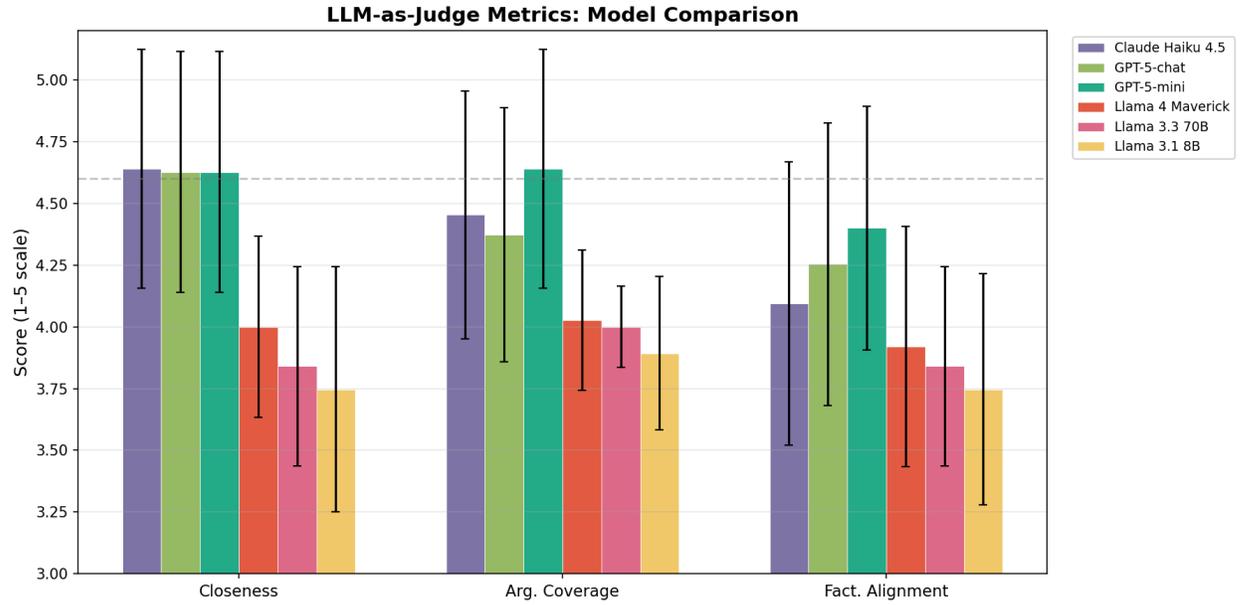

Figure 1: LLM-as-Judge scores across three evaluation dimensions. Error bars represent ±1 SD. The dashed line marks the Tier 1 threshold.

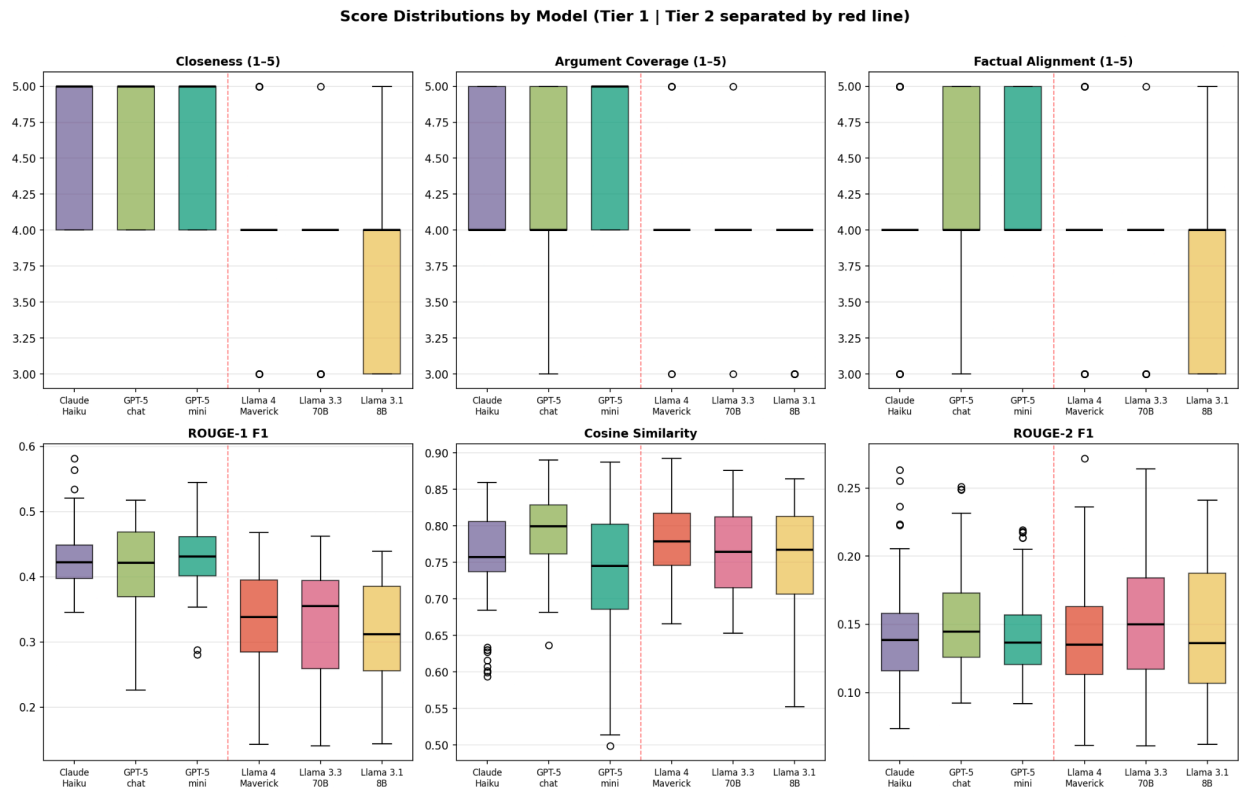

Figure 2: Score distributions across models for judge and programmatic metrics. Red dashed line separates Tier 1 (left) from Tier 2 (right).

## 5.2 The Efficient-Frontier Surprise: Claude Haiku Matches GPT-5

Claude Haiku 4.5, Anthropic's smallest model, is statistically indistinguishable from GPT-5 on overall closeness (p = 0.87). However, granular analysis reveals an asymmetric profile. On argument coverage, GPT-5-mini significantly outperforms Haiku (4.64 vs. 4.45, p = 0.022), and on factual alignment, Haiku falls behind both GPT-5 variants (4.09 vs. 4.25/4.40, p = 0.001 vs. GPT-5-mini). Haiku exhibits the largest coverage-to-factual gap ($\Delta = 0.36$): it identifies the right argument types but populates them with less accurate specifics—fabricated article numbers, incorrect dates, or hallucinated institutional references. This suggests strong structural legal reasoning but weaker domain-specific knowledge grounding.

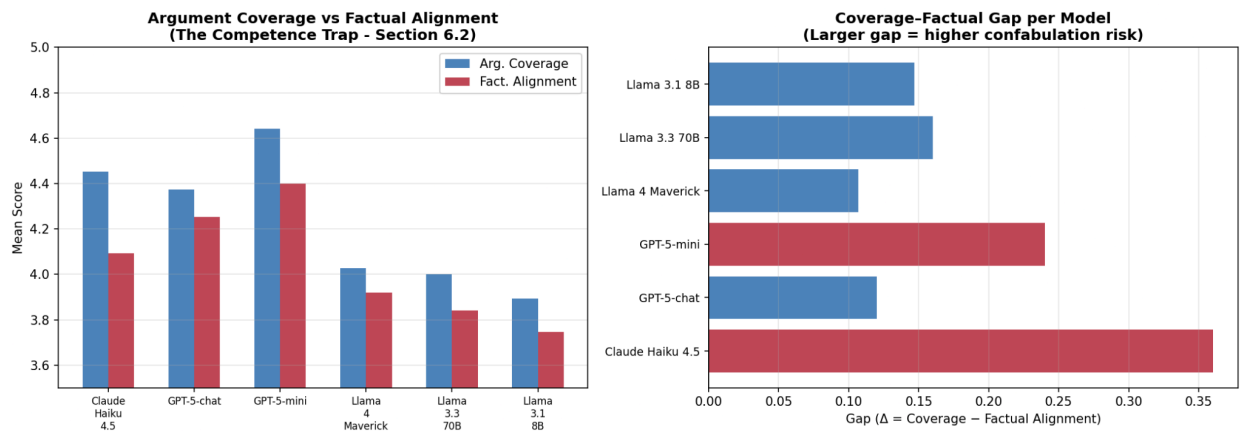

Figure 3: Argument Coverage versus Factual Alignment by model. Right panel shows the gap ($\Delta$ = Coverage − Alignment); larger gaps indicate higher confabulation risk.

## 5.3 Does Reasoning Help? GPT-5-mini vs. GPT-5-chat

GPT-5-mini (reasoning, built-in chain-of-thought) and GPT-5-chat (non-reasoning) are tied on closeness (both 4.63, p = 1.0). The reasoning advantage emerges only on argument coverage (4.64 vs. 4.37, p = 0.002): GPT-5-mini more completely enumerates justifications. It also trends toward better factual alignment (4.40 vs. 4.25, p = 0.14). Interestingly, GPT-5-mini scores the lowest cosine similarity (0.735) while GPT-5-chat scores the highest (0.792)—the first indication that embedding similarity and semantic quality diverge in this domain. Reasoning capabilities improve completeness but do not increase overall quality, suggesting the bottleneck is domain knowledge, not chain-of-thought depth.

### 5.4 The Open-Weight Plateau

Scaling from 8B to 70B parameters within the Llama family yields no significant improvement on closeness (3.75 vs. 3.84, p = 0.18) or factual alignment (p = 0.19). The only significant difference is a small gain in argument coverage (4.00 vs. 3.89, p = 0.01). This near-9× parameter increase produces a Cohen's d of only 0.41. Llama 4 Maverick (17B active / 400B total, MoE) scores significantly higher than Llama 3.1 8B (d = 0.58) but remains far below frontier models (d > 1.4), suggesting the between-tier gap is driven by training data composition and alignment rather than parameter scale.

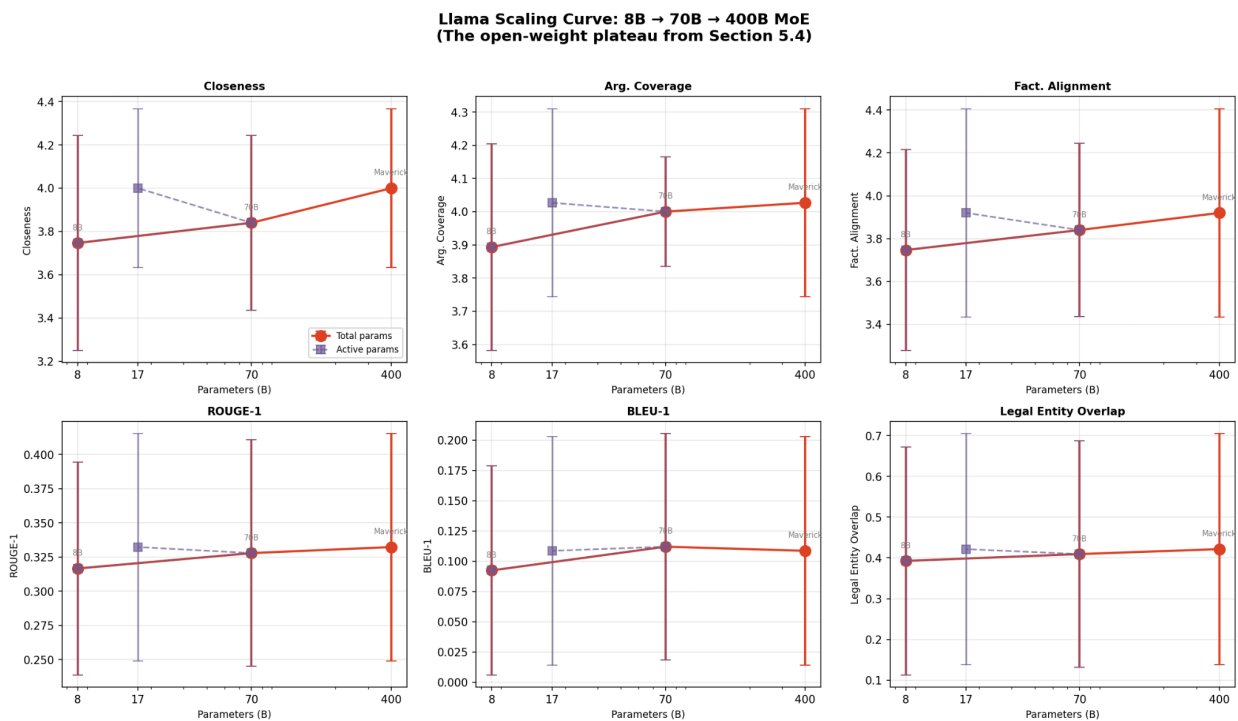

Figure 4: Llama family scaling curves across six metrics. The near-flat trajectory illustrates the open-weight plateau (Cohen's d = 0.41 for 8B→70B).

### 5.5 Failure Mode Analysis: What Makes a Law "Hard"?

Per-law analysis reveals systematic difficulty patterns. The easiest laws to predict were those implementing EU directives or making procedural amendments (mean closeness 4.53). The hardest were politically idiosyncratic proposals—such as earmarking funds for a specific religious institution (3.93) or niche regulatory modifications. This pattern suggests a predictability hierarchy: laws following well-known templates are predictable because their rationale follows recognizable structures, while laws driven by domestic political considerations require contextual knowledge LLMs are unlikely to possess. The difficulty ranking is largely consistent across tiers (per-law SD = 0.15–0.25), suggesting law-inherent complexity, not model idiosyncrasy, drives variance.

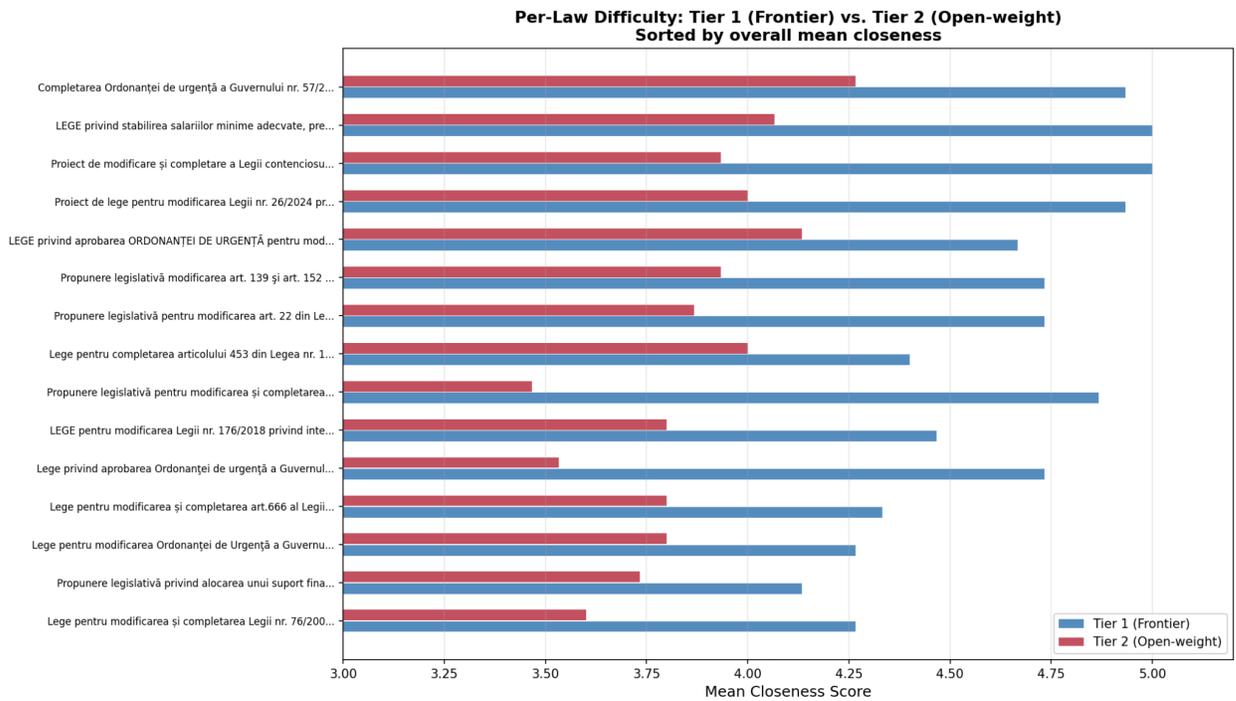

Figure 5: Perlaw closeness scores by tier. Laws sorted by overall mean difficulty (easiest at bottom). The gap between tiers is consistent across laws.

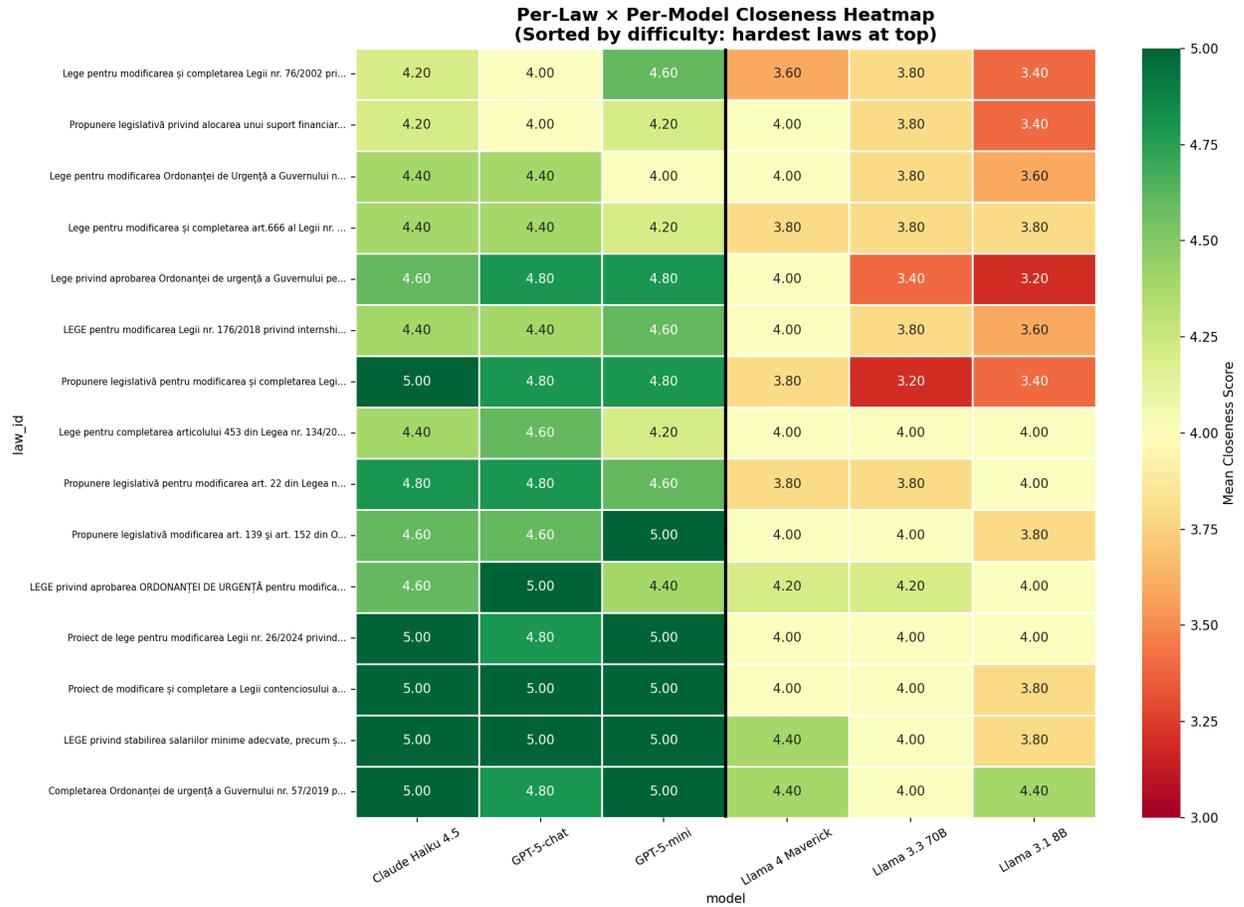

Figure 6: Per-law × per-model closeness heatmap (sorted by difficulty, hardest at top). Black line separates Tier 1 (left) from Tier 2 (right).

### 5.6 The Metric Divergence Problem

An interesting methodological finding is the near-complete dissociation between embedding-based similarity and semantic quality. Cosine similarity shows zero-to-negative correlation with judge closeness scores (r = −0.04 overall). Rank-level divergence is dramatic: cosine ranks GPT-5-chat first and GPT-5-mini last, while the judge ranks them identically. ROUGE-1 correlates moderately with judge closeness (r = 0.42), but ROUGE-2 and ROUGE-L fail to distinguish models (Kruskal–Wallis p = 0.24 and 0.18 respectively). These findings underscore the limitations of surface-level metrics for legal text evaluation, where equally valid rationales may use entirely different phrasing.

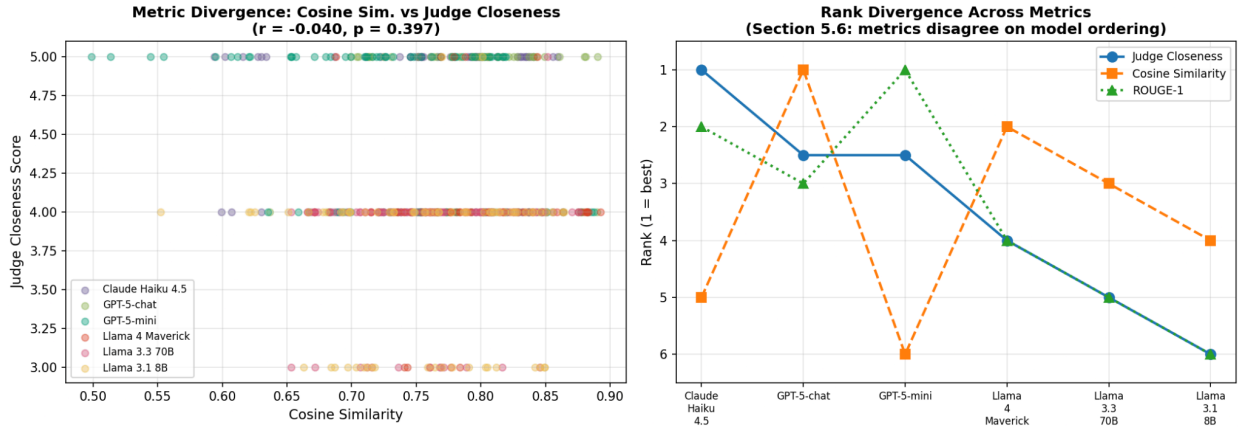

Figure 7: Metric divergence. Left: cosine similarity vs. judge closeness (r = −0.04). Right: rank disagreement across metrics — cosine ranks GPT-5-chat first and GPT-5-mini last, while the judge ranks them identically.

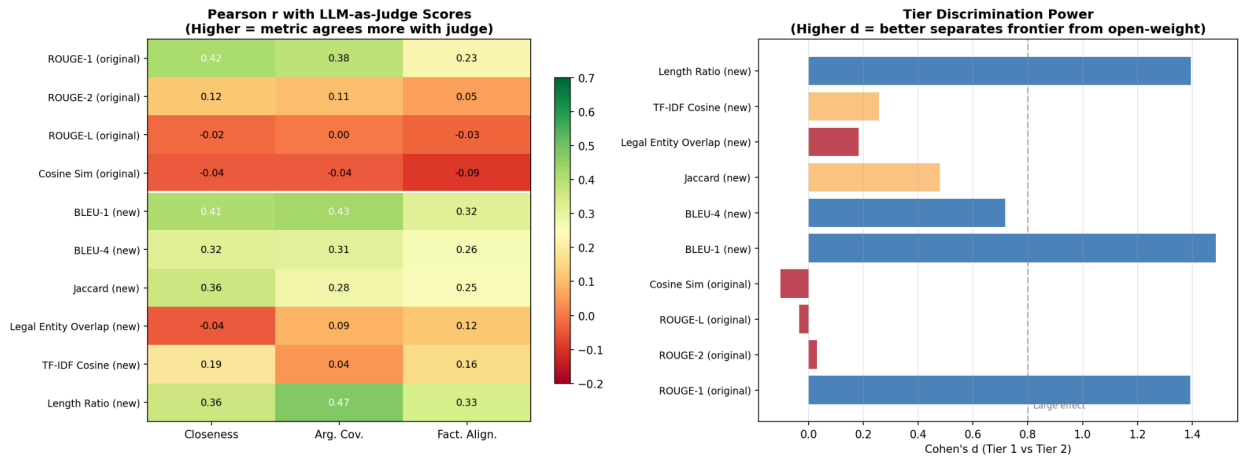

Figure 8: Metric reliability assessment. Left: Pearson correlation with judge scores. Right: tier discrimination power (Cohen's d). ROUGE-2 and ROUGE-L show nearzero discrimination

## 5.7 LLM-as-Judge Reliability

Several patterns in judge scoring warrant discussion. Score compression is evident in mid-tier models: Llama 3.3 70B receives exactly 4.0 on argument coverage in 97% of evaluations. The judge uses only the upper portion of its scale (minimum observed: 3.0), reducing discriminability. A response-length confound is substantial (r = 0.56 with closeness), with top-tier models generating longer responses (5,500–8,600 characters) compared to Tier 2

(2,900–3,200). Intra-model consistency ranges from SD = 0.23 (Llama 4 Maverick) to 0.37 (Llama 3.1 8B), suggesting adequate but imperfect reproducibility.

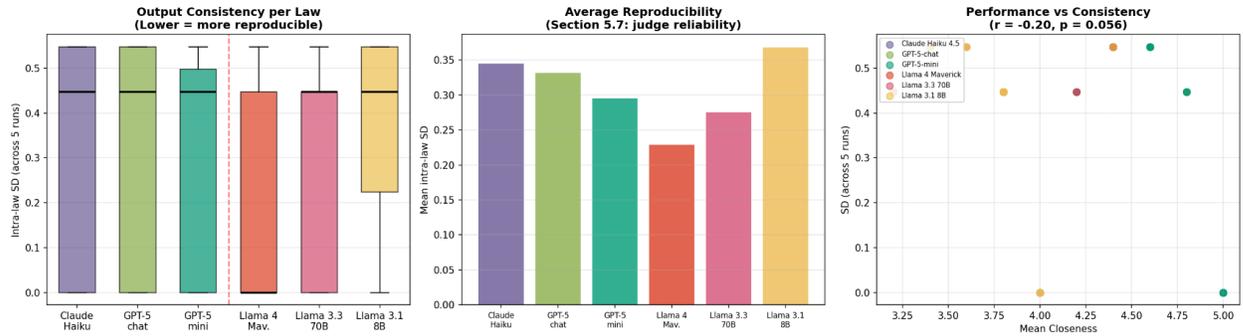

Figure 9: Intra-model consistency. Left: per-law standard deviation across 5 runs. Middle: mean SD per model. Right: performance vs. consistency.

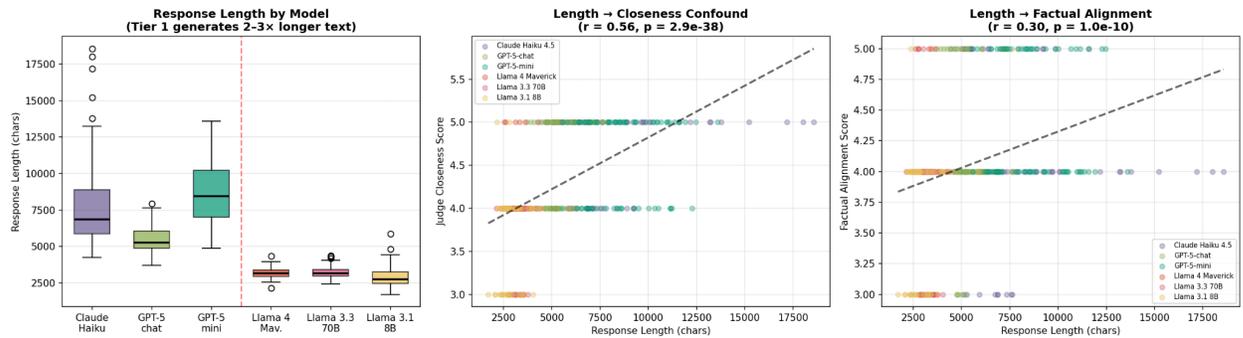

Figure 10: Response length confound. Tier 1 models generate substantially longer responses. Length correlates r = 0.56 with closeness, raising questions about judge verbosity bias.

# 6 Discussion

## 6.1 Adverse Selection and the Two-Tier Structure

The two-tier separation has direct consequences for the adverse selection problem (Eisenhardt, 1989). For LLMs in political advisory roles, no domain-specific quality signals exist. The top-tier models are not uniformly the most expensive or parameter-rich: Claude Haiku 4.5, a cost-optimized model, is indistinguishable from GPT-5 on aggregate closeness.

General-purpose benchmarks do not predict political advisory reliability. The information available in the market—pricing tiers, parameter counts, benchmark scores—is actively misleading for legislative advisory task selection. The within-tier compression reinforces this: scaling from 8B to 70B parameters produces only d = 0.41, suggesting the between-tier gap is driven by training data composition and alignment procedures rather than computational scale.

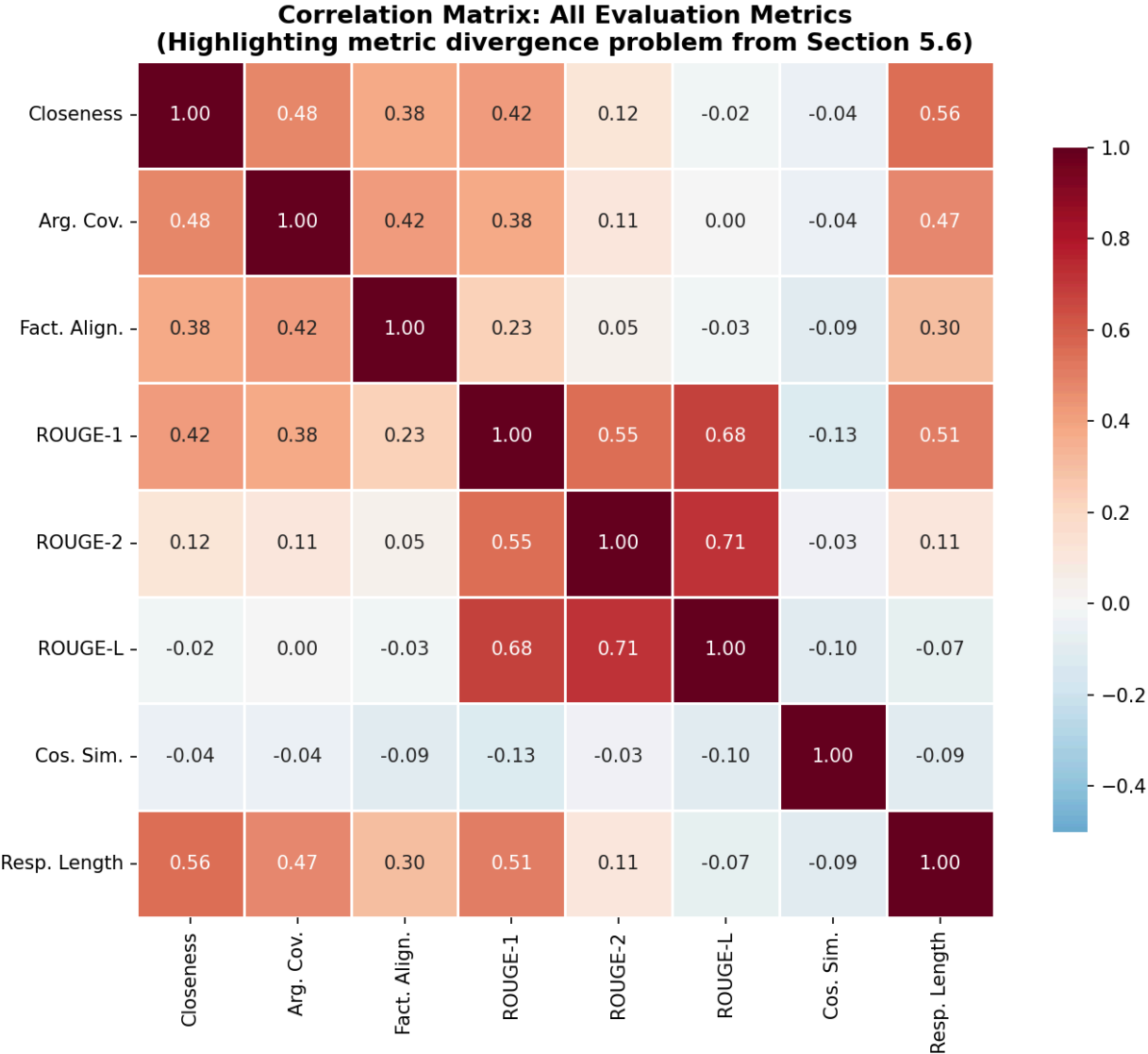

Figure 11: Correlation matrix across all evaluation metrics. Notable: cosine similarity shows near-zero correlation with judge closeness; response length correlates r = 0.56 with closeness.

## 6.2 The Competence Trap: Information Asymmetry After Selection

The most consequential finding is Claude Haiku's asymmetric profile: competitive on argument coverage but significantly weaker on factual alignment (Δ = 0.36). Within Simon's (1990) framework, a legislative staffer encounters output that identifies the right argument categories, uses appropriate terminology, and follows structural conventions of official memoranda. These surface features function as System 1 heuristic shortcuts (Kahneman, 2011): fluent outputs discouraging further scrutiny. The factual errors, fabricated article numbers, incorrect dates, are embedded within competent analytical structures and require domain expertise to detect. This creates a competence trap: rather than functioning as a bounded rationality mitigator, the model replicates the shortcuts Kahneman warns against while generating unwarranted confidence. This maps onto moral hazard (Eisenhardt, 1989), but the hazard is structural rather than behavioral, arising from statistical regularities rather than strategic misalignment.

### 6.3 The Predictability Hierarchy and Monitoring Limits

The cross-model judge successfully discriminates between tiers and identifies the factual alignment gap invisible to automated metrics. However, score compression, scale restriction, and the response-length confound constrain monitoring adequacy, instantiating the second-order agency problem identified in Section 3.3. The metric divergence problem (r = −0.04 between cosine similarity and judge scores) means that computationally inexpensive monitoring proxies will yield systematically incorrect quality assessments. The monitoring investment required for political advisory evaluation is substantially higher than for other LLM applications.

The per-law predictability hierarchy clarifies where monitoring costs are justified. For EU directive transpositions, LLMs function as bounded rationality mitigators, organizing reasoning the legislator could find but would require time to assemble. For politically idiosyncratic proposals, models generate plausible rationales from general templates rather than actual political motivations. This aligns with Gigerenzer and Gaissmaier's (2011) argument that heuristic adequacy depends on environmental structure, and extends Yuksel et al.'s (2025) finding about language-dependent performance to the broader problem of training data representation for non-Anglophone legislative systems.

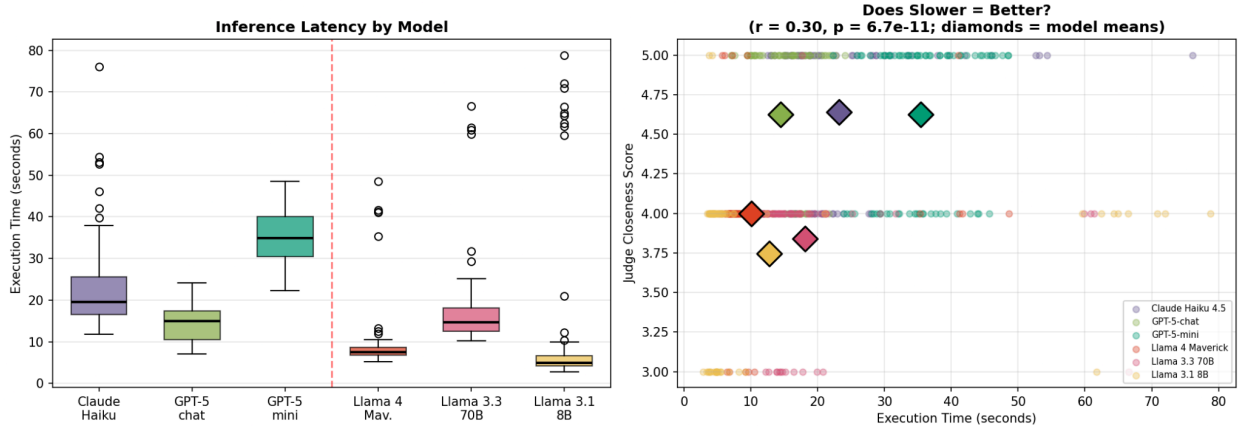

Figure 12: Inference latency versus quality. Diamond markers show per-model means. Slower inference does not guarantee higher quality.

### 6.4 Synthesis

The findings reveal that the three layers of our framework, bounded principals creating demand, bounded LLM agents generating supply, and bounded monitors evaluating quality, interact in ways that compound failures. The legislator-principal is bounded by time and processing capacity (Simon, 1990). The LLM-agent is bounded by training data coverage, encoded patterns, and RLHF objectives. The monitor captures structural quality but misses factual precision. A hallucination by the LLM-agent is more likely to survive monitoring by a judge lacking domain knowledge and more likely to influence a principal lacking time for verification. Conversely, when the task environment is well-structured and well-represented in training data, all layers perform adequately and the advisory relationship delivers genuine value.

This reframes how the political bias literature conceptualizes LLM risk. The dominant framing treats bias as a fixed ideological property (Rozado, 2024; Yang et al., 2024). Our results suggest that in applied contexts, the operative risk is task-dependent confabulation shaped by training data coverage. A model that performs neutrally on abstract quizzes may produce misleading advisory output for Romanian legislation, not from ideological agenda but from contextual ignorance. Furthermore, in Jensen and Meckling's (1976) original formulation, misalignment arises from divergent interests amenable to incentive correction. An LLM has no interests in the relevant sense; its misalignment arises from structural construction properties.

The agency cost is not correctable through contractual mechanisms but must be managed through task selection, institutional design pairing LLM output with human verification, and monitoring investment calibrated to the predictability hierarchy.

## 7 Conclusion

This study asked how reliable commercial LLMs are as political advisory tools when evaluated against official legislative reasoning from the Romanian Senate. The answer is conditional: reliability varies systematically with the interaction between model capability, task structure, and monitoring design. Frontier models achieve mean semantic closeness above 4.6/5.0 for Romanian law proposals, while open-weight models cluster a full tier below (d > 1.4). The advisory output of frontier LLMs is substantially aligned with official reasoning in aggregate, but this masks consequential variation across task types.

Addressing RQ1, frontier models are statistically indistinguishable on overall closeness despite substantial differences in architecture and pricing. Claude Haiku 4.5 matches GPT-5 holistically while exhibiting the largest factual alignment gap. GPT-5-mini's reasoning capabilities improve argument coverage but not overall quality, suggesting domain knowledge rather than chain-of-thought depth is the bottleneck. Within the open-weight family, scaling from 8B to 70B parameters yields marginal gains (d = 0.41), indicating the performance gap is driven by training data composition rather than parameter scale.

Addressing RQ2, the form of bias most relevant to applied legislative advisory is not the stable ideological lean documented through abstract quizzes but task-dependent confabulation shaped by training data coverage. All models perform adequately on well-templated legislative proposals and struggle with politically idiosyncratic ones. The operative risk for legislators is contextual ignorance: models generating plausible but unfounded reasoning for tasks outside their training distribution. This reframes the political bias question from an ideological problem to an epistemological one.

The study contributes the first practitioner-oriented evaluation using real parliamentary data, extends the bias literature to applied legislative contexts, and proposes cascading bounded rationality as a framework for human-AI advisory relationships where failures compound across bounded principals, agents, and evaluators. Limitations include the 15-proposal dataset size, score compression in the LLM-as-Judge, the response-length confound, and single-timepoint evaluation. Future work should expand across EU member states and languages, track performance longitudinally, test the competence trap experimentally with legislative staffers, and develop domain-specific legal text evaluation metrics.